\documentclass[amssymb]{article}

\usepackage{amssymb,amsmath,verbatim}
\usepackage{graphicx}
\usepackage{epsfig}
\usepackage{dcolumn}
\usepackage{bm}
\usepackage{authblk}

\newcommand{\be}{\begin{equation}}
\newcommand{\ee}{\end{equation}\noindent}
\newcommand{\eei}{\end{equation}}
\newcommand{\bea}{\begin{eqnarray}}
\newcommand{\eea}{\end{eqnarray}\noindent}
\newcommand{\eeai}{\end{eqnarray}}

\newcommand{\hf} {\frac{1}{2}}
\newcommand{\nn}{\nonumber\\}

\newcommand{\eq}[1]{(\ref{#1})}
\newcommand{\ord}[1] {{\cal O}(#1)}
\newcommand{\C}[1]{{\cal#1}}
\newcommand{\B}[1]{{\bar#1}}

\begin{document}
\title{Interpolation formulas for asymptotically safe cosmology}

\author[1]{S. Nagy}
\affil[1]{Department of Theoretical Physics, University of Debrecen,
P.O. Box 5, H-4010 Debrecen, Hungary}
\author[2]{K. Sailer}
\affil[2]{Department of Theoretical Physics, University of Debrecen,
P.O. Box 5, H-4010 Debrecen, Hungary}

\date{\today}
\maketitle
\begin{abstract}
Simple interpolation formulas are proposed for the description of the renormalization group (RG) scale dependences of the gravitational couplings in the framework of the  2-parameters Einstein-Hilbert (EH) theory of gravity and applied to a simple, analytically solvable, spatially homogeneous and isotropic, spatially flat model universe.
The analytical solution is found in two schemes incorporating different methods of the determination of the 
 conversion rule $k(t)$ of the RG scale $k$ to the cosmological time $t$. In the case of the discussed model these schemes turn out to yield identical cosmological evolution.
  Explicit analytical formulas are found for the  conversion rule $k(t)$ as well as for the characteristic time scales $t_G$ and $t_\Lambda>t_G$ corresponding to the dynamical energy scales $k_G$ and $k_\Lambda$, respectively, arising form the RG analysis of the EH theory. It is shown that there exists a model-dependent time scale $t_d$ ($t_G\le t_d<t_\Lambda$)  at which the accelerating expansion changes to the decelerating one. It is shown that the evolution runs from a well-identified cosmological fixed point to another one. As a by-product we show that the entropy of the system decreases monotonically in the interval $0<t\le t_\Lambda$ due to the quantum effects.
\end{abstract}


\section{Introduction}

Asymptotically safe cosmology \cite{Litim2011,BonSau2017,Mandal2021,TyeXu2010,Weinberg2010,BieSau2016} relies on the success of the asymptotic safety scenario \cite{Weinberg1977} in quantum gravity achieved in the last two decades (see the status reports in \cite{ReuterSau2012,Reuter:2019byg,BonSau2020} and the references therein). 
Based on the RG studies of the 4-dimensional 2-parameters EH gravity in the continuum a good evidence is found on the existence of an ultraviolet (UV) fixed point, the so-called Reuter fixed point  and that of the Gaussian fixed point \cite{Reuter1996,LauReu2001a,LauReu2001b,Bonanno:2004sy,Litim2004,ReuterSau2007,Codello:2008vh,Christiansen:2012rx,Dona:2013qba,Dupuis:2020fhh,Trivedi2022}. It is assumed that the early time evolution of the Universe is governed by the Reuter fixed point revealing asymptotic safety, while its  evolution preceding somewhat the present day time  is dictated by the perturbative regime of the gravitational couplings near the Gaussian fixed point. The idea is  that at any cosmological the time $t$ the relevant physical processes in the  Universe are those of a given energy scale $k=k(t)$ identified with the RG scale $k$. The function $k(t)$ is called below the $k$-to-$t$ conversion rule. As the authors pointed out in \cite{Saueressig2019} the main features of the RG flow of the gravitational couplings  are rather independent of the fine details of RG scheme used as well the matter content of the universe. Moreover, the RG analyses of the Euclidean and Lorentzian theories yield rather similar results in that respect \cite{Manrique:2011jc,Pawlowski:2015mia,Nagy:2019jef,Knorr:2021niv,Platania:2022gtt}. These features of the RG flow on realistic RG trajectories include {\em (i)} 3 scaling regions, the UV regime governed by the Reuter fixed point, the crossover regime ended in the perturbative one close to the Gaussian fixed point, and the IR region where the gravitational couplings become almost scale-independent, and {\em (ii)} the order of magnitude of their limit points at $k_G\sim \ord{m_{\rm{Pl}}}$ and $k_\Lambda\sim \ord{10^{-30}k_G}$, where $m_{\rm{Pl}}$ is the Planck mass. For the expanding Universe one naturally expects that decreasing energy scale $k$ should correspond to increasing cosmological time  $t$. The RG studies prove that Newton's gravitational coupling $G$ vanishes at the Reuter fixed point, making plausible that the quantum improved  evolution equations, the quantum improved Friedmann equations   should keep their classical form  with the only modification that Newton's gravitational constant $G_0$ and the cosmological constant $\Lambda_0$ (the null index indicates the present day values) should be replaced by the time-dependent couplings $G(t)\equiv G(k(t))$ and $\Lambda(t)\equiv \Lambda(k(t))$, respectively. From now on we assume that the functions $G(k)$ and $\Lambda(k)$ are well-known functions from the RG analysis of quantum gravity.

So far quantum fluctuations of the metric and those   of the matter fields are neglected, the cosmological evolution of the homogeneous and isotropic Universe is described by the quantum modified  Friedmann equations and their consistency condition, where the gravitational constants are replaced by their time-dependent counterparts. These equations represent the symmetry reduced sector 
of the quantum improved version of classical EH gravity, when the gravitational constants $G_0$ and $\Lambda_0$ are replaced by their  time-dependent counterparts. Then the Bianchi identity  $\nabla_\mu G^{\mu\nu}=0$ (for the Einstein tensor $G^{\mu\nu}$) implies the quantum improved consistency condition $\nabla_\mu\lbrack G(t)T^{\mu\nu}\rbrack=0$ of the Einstein equations with the stress-energy tensor of matter $T^{\mu\nu}$. Having done the symmetry reduction to the homogeneous and isotropic sector, one gets the quantum improved consistency condition (c.f. \eq{consis} below) of the quantum improved Friedmann equations, which is now different   of the law of the  local energy conservation of matter, as it was in the classical case.
 Let say that one formulates the cosmological evolution problem in terms of the Hubble parameter $H(t)$, the energy density of matter $\rho^{(m)}(t)$, and the function $k(t)$. Then one has only 2 independent equations for the determination of 3 yet unknown functions. Regarding this problem, various approaches have been worked out in the literature. The first attempts assumed that one
has to make some intuitive assumption on the function $k(t)$ like $k=\xi/t$ with some constant $\xi$, require the local conservation of the energy  of matter separately, like in classical cosmology \cite{Weinberg}, and adjust the constant $\xi$ in the UV and in the perturbative regimes separately in order to achieve consistency among the 4 equations \cite{ReuBon2001,BonReu2002,BonannoReu2002}. These first efforts have made a hint on the scale- or time-dependence of $\xi$ itself: different results have been obtained for $\xi$  in the UV and in the perturbative regimes. There is an approach 
that determines the function $k(t)$ from the interplay between  the local energy  conservation  of matter 
 and  the reduced consistency condition (c.f. \eq{rconsis} below) following from the quantum improved consistency condition  \cite{Babic2005}. Below we shall follow this route under Scheme A.
Another approach has been proposed in the framework of dimensionless cosmological variables \cite{Copeland1998,Copeland2006}, when the $k$-to-$t$ conversion rule is determined from a constraint on the RG parameters (c.f. \eq{conqi} below) which follows from taking the first derivative of the Friedmann constraint with respect to the RG scale $k$ \cite{Litim2011}. 
This approach is particularly adequate to consider models in which both gravitation and matter underly quantum effects, but we restrict now ourselves to a model, where  matter is represented by a barotropic fluid with classical equation of state (EOS).  
Below we shall follow this route under Scheme B.  
 Finally, we have to mention the approach when one  gives up the requirement of  the local energy conservation of matter separately, then one has to go back to some intuitive choice of the conversion rule $k$-to-$t$, and establishes that matter is  expanding in a nonadiabatic manner, and the entropy production can be read off from the thermodynamical reinterpretation of the quantum improved consistency condition \cite{Mandal2021}. We should mention, that in particular systems both Scheme A and Scheme B may 
 reproduce the naive inverse proportionality $k\propto 1/t$, but not in general.   It has been shown that according to the matter content of the Universe, the cosmological evolution may exhibit  cosmological fixed points, where the RG scale freezes in and ones, where the RG scale continues to evolve; a detailed classification of the possible cosmological fixed points can be found in \cite{Litim2011}.

The main goal of the present paper is to put forward interpolation formulas that describe the above mentioned main features of the RG flow of the gravitational couplings. Then we shall apply these formulas to a particularly simple, analytically solvable model universe,  in order to obtain analytic relations for the characteristic time scales of the cosmological evolution. We restrict ourselves to the asymptotically safe cosmology based on the 2-parameters EH gravity. We intend to give on interpolation formulas which reproduce the main features of the RG flow discussed in details in \cite{Saueressig2019}. 
 The UV scaling  for  $k>k_G$ is governed by the Reuter fixed point, while for $0<k\le k_\Lambda$ the gravitational couplings take their present day  constant values $G_0=m_{\rm{Pl}}^{-2}$ measured at the laboratory scale $k_l\approx 8.2\times 10^{-34}m_{\rm{Pl}}\approx 10^{-3}k_\Lambda$ and $\Lambda_0 \approx 2.7\times 10^{-122} m_{\rm{Pl}}^2$ observed at the Hubble scale $k_H\approx  8.2\times 10^{-62} m_{\rm{Pl}}\ll k_\Lambda$. (It should be noticed that along the RG trajectories relevant for the evolution of our Universe there may exist a deep IR regime for $k\lesssim k_l\sim \ord{10^{-3} k_\Lambda}$, where the gravitational couplings $G$ and $\Lambda$ tend to zero in the limit $k\to 0$ \cite{BieSau2016}; this scaling region can affect, however, the evolution of the Universe in the late future, which is out of the scope of our discussion in the present paper.)  

On the space of the dimensionless couplings $(g=G k^2,~\lambda= \Lambda/k^2)$ the physical RG trajectory relevant for our Universe starts at the Reuter fixed point $(g_*\approx 0.707,\lambda_*\approx 0.193)$ \cite{fnote1},  for $k\approx k_\Lambda$ it approaches the Gaussian fixed point at $(g_G=0, \lambda_G=0)$, and with the further decrease of the RG scale $k$ runs away from the Gaussian fixed point towards positive values of $\lambda$, while the dimensionful couplings take their constant values observed in the present day.  The scaling of the couplings slightly above the scale $k_\Lambda$ is the so-called perturbative regime. The proposed interpolation formulas recover the UV scaling laws, the constant values  $G_0$ and $\Lambda_0$ below the scale $k_\Lambda$, 
 and in the crossover regime $k\in [k_\Lambda,k_G]$ are motivated by the scaling in the perturbative regime. The interpolation formulas contain 3 free parameters which are determined from the continuity of $G(k)$, $\Lambda(k)$ and that of the matter density $\rho^{(m)}$ at the dynamical scale $k_G$, while the scale $k_\Lambda\approx 8.2\times 10^{-31} m_{\rm{Pl}}$ is taken from the RG analyses \cite{Saueressig2019}.

  The interpolation formulas are applied to a rather simple model universe making the assumptions that {\em (i)} the universe is spatially homogeneous and isotropic, {\em (ii)} it is spatially flat,  {\em (iii)} its matter content is assigned to a single type of barotropic fluid with the EOS $p^{(f)}=w\rho^{(f)}$ with the constant  $0\le w<1$ and  $p^{(f)}$ and $\rho^{(f)}$ being its pressure and energy density, respectively, and {\em (iv)} the barotropic fluid is subjected to the law of local energy conservation separately. In this model we identify the presence of the evolving cosmological coupling $\Lambda$ with that of the  dark energy and call the quantity 
\bea\label{rhoLamdf}
\rho^{(\Lambda)}&=&\frac{\Lambda}{8\pi G}
\eea
  the density of the dark energy. It is rather giving a name to the $\Lambda$-component of the model; to go beyond the nature of dark energy is out of the scope of the present paper. Motivated by the interpretation used in \cite{Litim2011} we take the point of view  that $\rho^{(\Lambda)}$   can be considered as the field-independent potential energy density of a condensed scalar field, which implies the EOS $p^{(\Lambda)}=-\rho^{(\Lambda)}$ with the pressure $p^{(\Lambda)}$ of the dark energy.

It is well-known  that if  the universe is filled by  radiation or dust  after the Big Bang,  the quantum improvement may make it unnecessary any fine-tuning of  the initial conditions or the  introduction of  an inflaton field, in order to solve the horizon and flatness problems  \cite{ReuBon2001,BonannoReu2002}. In that case the physical phase trajectory runs to a universal attractor, being independent of the IR values of the gravitational couplings. The early time evolution is characterized by the time-dependent Hubble parameter $H=h/t$ and the scale factor $a(t)=\C{M} t^h$ with $h=\frac{4}{3(1+w)}$. For $h\ge 1$ both the flatness and  horizon problems  are solved automatically \cite{ReuBon2001,BonannoReu2002}: {\em (i)}  the universal attractor starts from a singularity at $t=0$, characterized by constant and equal cosmological parameters  $\Omega^{(m)}$ and $\Omega^{(\Lambda)}$ of the matter and that of the dark energy, respectively, so that $\Omega^{(m)} =\Omega^{(\Lambda)}=\hf$ implies in total $\Omega^{(tot)}=1$ and  flat spatial geometry; {\em( ii)}  no finite particle horizon exists, so that  our entire visible universe should have been causally connected in the Planck era. 

  Applying the proposed interpolation formulas to the above described spatially flat  model universe, enables one to solve 
 analytically the problem of evolution  for times from the Big Bang at $t=0$ to the asymptotically far future $t\to \infty$. 
 This circumstance enables one to obtain analytic expressions for the  $k$-to-$t$ conversion rule in the Planck and  crossover eras as well as those for the characteristic time scales $t_G$ and $t_\Lambda$ corresponding to the dynamical RG scales $k_G$ and $k_\Lambda$, respectively.  The discussion of the evolution problem is performed both in Scheme A and  Scheme B.
Both descriptions provide identical results for the entire cosmological evolution including the characteristic time scales.
This is basically the consequence of the equivalence of the reduced consistency condition and the constraint on the RG parameters, eqs. \eq{rconsis} and \eq{conqi}, respectively, in the case of the discussed particular model universe. After continuously sewing the solutions on the various intervals for $k(t)$ at $t=t_G$ and those for $H(t)$ at $t=t_\Lambda$ 
 the cosmological quantities  $H(t)$, $\rho^{(f)}(t)$, and $\rho^{(\Lambda)}(t)$ become continuous.  In the Planck era for $0<t\le t_G$ the well-known universal evolution governed by the Reuter fixed point is recovered with  the simple conversion rule $k\propto 1/t$, while for  late times $t\ge t_\Lambda$ the classical evolution holds trivially. In the crossover era for $t_G\le t\le t_\Lambda$,  a nontrivial, with increasing cosmological time strictly monotonically decreasing function $k(t)$ is obtained. Nevertheless, the simple proportionality $ t_\Lambda\propto k_\Lambda$ remains yet a rather good   approximation. The solution found in Scheme B enables us to identify the  evolution in the time interval $0<t\le t_G$ as a  cosmological fixed point with evolving RG scale $k(t)$  and the dominance of a   mixture of dark energy and barotropic fluid in equal proportions. The evolution of the model universe  in  the late future  for $t>t_\Lambda$ corresponds to another cosmological fixed point with freezed in RG scale at $k_\Lambda$ and dark energy domination in the asymptotic limit $t\to \infty$. It  is  argued that the evolution of dark energy is accompanied with the decrease of its entropy in the Planck and crossover eras, while the entropy of the barotropic fluid is preserved.

\section{Interpolation formulas}

In this section, we propose simple  formulas  reflecting the main features of the RG flow of  the gravitational couplings $G$ and $\Lambda$ in the 4-dimensional, 2-parameters EH gravity and interpolating  between   the UV and  IR scaling regimes. 
In the parameter space of the dimensionless couplings $(g,~\lambda)$ 
RG trajectories are considered along which both of the couplings are nonnegative, running in the broken symmetry phase.  The physically relevant trajectories  emanate from the  Reuter fixed point $(g_*,~\lambda_*)$, pass  the close neighbourhood of the Gaussian fixed point, and possibly approach an IR fixed point for $k\to 0$. The latter is expressed in our treatment by saying that the couplings $G$ and $\Lambda$ take their present day observed values in the  IR region. This behaviour,  
 based on the results of much efforts on the RG analysis of the EH-theory of general relativity, is  thoroughly discussed in \cite{Saueressig2019}. There are two dynamical scales characterizing the RG flow: the scale $k_G$ above which the UV scaling laws $G=g_*/k^2$ and $\Lambda=\lambda_*k^2$ hold and the scale $k_\Lambda$ below which the  IR regime with $G=G_0$ and $\Lambda=\Lambda_0$ is found. At the scale $k_\Lambda$ the RG trajectory is close to the Gaussian fixed point, so that for scales $k$ larger but close to $k_\Lambda$ the perturbative scaling laws can be used (see,  e.g.  eqs. (1.2), (1.3) in \cite{Mandal2021} or eq. (26) in \cite{Saueressig2019}).  These perturbative laws motivate our choice of the interpolation formulas in the crossover regime $k\in [k_\Lambda, k_G]$. Thus the following simple formulas are proposed,
    \bea\label{Gk}
  G(k)&=& \biggl\{ \begin{array}{ccc} G_0 &{\rm{for}} & 0\le  k\le k_\Lambda\cr
                              G_0-b(k^2-k_\Lambda^{2}) &{\rm{for}} &k_\Lambda< k<k_G\cr
                                   g_*k^{-2} & {\rm{for}} & k_G\le k  \end{array},\\
\label{Lambdak}
 \Lambda(k) &=&\Biggl\{ \begin{array}{ccc} 
   \Lambda_0 & {\rm{for}} & 0\le k<k_\Lambda \cr
    \Lambda_0 + c (k^4-k_\Lambda^4)  &  {\rm{for}} & k_\Lambda\le k<k_G\cr
    \lambda_* k^2 & {\rm{for}} & k_G\le k
\end{array},
\eea
where $k_\Lambda\approx 8.2\times 10^{-31} m_{\rm{Pl}}$ is taken from the RG analysis \cite{Saueressig2019} and $k_G, ~b,~c$ are yet  free parameters. By construction, the formulas are  continuous at the scale $k=k_\Lambda$. Two algebraic equations for the determination of the 3  yet unknown parameters are the  continuity conditions for $G$ and $\Lambda$ at the scale $k_G$.  The third equation is obtained by requiring the continuity of the ratio 
of the time derivatives of the gravitational couplings $r_{{\dot \Lambda}/{\dot G}}\equiv \frac{d\Lambda(k)/dk}{dG/dk}$ at the scale $k_G$. (Here and below the dot denotes the  time derivative.) Since local energy conservation is supposed to hold separately for matter, the consistency condition $\nabla_\mu(GT^{\mu\nu})=0$ reduces to an expression, the reduced consistency condition (c.f. eq. \eq{rconsis} below) according to which the energy density of matter  is proportional to the ratio $r_{{\dot \Lambda}/{\dot G}}$. Therefore the requirement of the continuity of that ratio ensures the continuity of matter density  $\rho^{(m)}$ at the scale $k_G$. 
  We shall see that the three requirements settle $k_G$ at a value of the correct order of magnitude, $k_G\sim\ord{m_{Pl}}$. 
The continuity conditions are then the following,
\bea\label{c1}
  \Lambda(k_G) &=&\Lambda_0 + c(k_G^4-k_\Lambda^4)= \lambda_* k_G^2,\\
\label{c2}
G(k_G) &=& G_0 -b(k_G^2-k_\Lambda^2) =\frac{g_*}{k_G^2},\\
\label{c3}
r_{{\dot\Lambda}/{\dot G}}(k_G)&=& -\frac{2ck_G^2}{b} = -\frac{\lambda_*}{g_*}k_G^4.
\eea
Making the working hypotheses that there hold the order-of-magnitude estimates $k_G\sim m_{Pl} \gg k_\Lambda$ and $ck_G^4 \gg \Lambda_0$, one obtains for the free parameters of the interpolation formulas that
\bea\label{params}
&& k_G^2=\frac{3g_*}{G_0}, ~~b=\frac{2}{9g_*}G_0^2,~~c=\frac{\lambda_*}{3g_*} G_0.
\eea
Then we see that  our working hypotheses are rather  well satisfied, because the following estimates can be done:  $k_G=\sqrt{3g_*} m_{\rm{Pl}}\approx 1.46 m_{\rm{Pl}}$ and  $ck_G^4=\lambda_*k_G^2 =\frac{3g_*\lambda_*}{G_0}
\approx  0.4m_{\rm{Pl}}^2 \gg \Lambda_0 \approx 3\cdot 10^{-122} m_{\rm{Pl}}^2$.

One can check that the interpolation formula \eq{Gk} provides a good approximation of the IR scaling law for scales close to but above the scale $k_\Lambda$  given by  eq. (26) in \cite{Saueressig2019}. The latter can be approximated for $k\gtrsim k_\Lambda$, i.e. for $k-k_\Lambda\ll k_G$  by
\bea\label{IRGk}
 G(k) &\approx &G_0 \biggl( 1 - \frac{7}{6\pi}G_0(k^2-k_\Lambda^2) \biggr),
\eea
where $G(k_\Lambda)=G_0$ has been used. A comparison of the estimate \eq{IRGk} with our interpolation formula \eq{Gk} yields $b^{(est)} =\frac{7}{6\pi}G_0^2\approx 0.37 G_0^2$, while  according to our setting above $b=\frac{2}{9g_*} G_0^2\approx 0.31G_0^2$.

Finally it has been checked numerically that the interpolation formulas \eq{Gk} and \eq{Lambdak} with the parameter setting \eq{params} reproduce fairly well the RG evolution of the couplings $G$ and $\Lambda$ over the entire range of the RG scale covering roughly 40 orders of magnitude (see Fig. 4 in \cite{Saueressig2019}).

\section{Model universe}

In this section, we shortly overview the equations governing the evolution of the model universe with the properties
 {\em (i)-(iv)}. Below we intend to follow the evolution of this simple model in Schemes A and B independently.

\subsection{Scheme A} 

In order to be somewhat more general, let us consider models with the properties {\em (i), (ii)} containing  some matter  with energy density $\rho^{(m)}$ and pressure $p^{(m)}$. Assuming adiabatic expansion of matter implies the law of local energy conservation
\bea\label{encon}
 {\dot \rho}^{(m)} + 3H({ \rho}^{(m)}+p^{(m)}) &=&0,
\eea
where $H={\dot a}/a$ is the Hubble parameter and $a$ denotes the scale factor of the universe. 
The local energy conservation for matter  does not involve  the gravitational couplings explicitly, so that it keeps its nonimproved form. Nevertheless, the time-dependence of the Hubble parameter $H$ is affected by quantum effects.
The quantum-improved  Friedmann equations are given via the replacement of the observed constants $G_0$ and $\Lambda_0$ by their time-dependent counterparts $G(t)$ and $\Lambda(t)$:
\bea\label{Fr1}
  H^2 &=&\frac{8\pi G(t)}{3} \rho(t)= \frac{8\pi G(t)}{3} \rho^{(m)} +\frac{\Lambda (t)}{3} ,\\
\label{Fr2}
 \frac{\ddot a}{a} &=& - \frac{4\pi G(t)}{3} (\rho^{(m)}+3p^{(m)}) +\frac{\Lambda (t)}{3}.
\eea
Eq. \eq{Fr2} can be replaced by the more simple equation
\bea\label{Fr3}
{\dot H} &=& -4\pi G(t) (\rho+p)=-4\pi G(t)(\rho^{(m)}+p^{(m)}),
\eea
where $\rho=\rho^{(m)}+\rho^{(\Lambda)}$ and $p=p^{(m)}+p^{(\Lambda)}$ stand for the total energy density and pressure, respectively. The critical density $\rho_c=H^2/(8\pi G(t))$ is identical with the total energy density, since we restricted ourselves to models with flat spatial geometry. 
Eqs. \eq{Fr2} and \eq{Fr3} are dynamical equations, whereas eq. \eq{Fr1} is the so-called Friedmann constraint laid on the cosmological variables.  Eqs. \eq{Fr1} and \eq{Fr2} imply the consistency condition
\bea\label{consis}
\!\!\!\!\!\!\!\! \!\! {\dot \rho}^{(m)}+ 3H( \rho^{(m)}+p^{(m)}) &=& - \frac{ 8\pi \rho^{(m)}{\dot G} +{\dot \Lambda} }{8\pi G}
\equiv \C{P}^{(m)},
\eea
which is the symmetry reduced form of the condition $\nabla_\mu( GT^{\mu\nu})=0$.
Combining the latter with the local energy conservation for  matter, one finds the reduced consistency condition
\bea\label{rconsis}
 8\pi\rho^{(m)} {\dot G} +{\dot \Lambda } &=&0,
\eea
i.e., the vanishing of the quantity $\C{P}^{(m)}$. The condition $\C{P}^{(m)}=0$ expresses that during the evolution the entropy of the matter is conserved \cite{Mandal2021}. Assuming the consistency condition \eq{consis} as well as the law of local energy conservation of matter is in agreement with the treatment used in \cite{Weinberg,ReuBon2001,BonReu2002}.

 Although our interpolation formulas are not differentiable at $t=t_G$ and $t=t_\Lambda$, the continuity of the matter density $\rho^{(m)}$ ensures the continuity of the Hubble parameter $H$ and that of its first time derivative via the  Friedmann equations   \eq{Fr1} and \eq{Fr3}, respectively, which implies the continuity of the scale factor $a$ and its first and second time derivatives. Furthermore, continuity of the matter density $\rho^{(m)}$ implies that of its first time derivative through eq. \eq{consis} with $\C{P}^{(m)}=0$.

It is noteworthy to make here a note on the entropy of the dark energy. In the case of time-dependent $G$ and $\Lambda$ it holds the equation
\bea\label{PLambda1}
 {\dot \rho}^{(\Lambda)}+3H(\rho^{(\Lambda)}+p^{(\Lambda)})&=& {\dot \rho}^{(\Lambda)}=\C{P}^{(\Lambda)}
\eea
with 
\bea\label{PLambda2}
\C{P}^{(\Lambda)}&=& \frac{  {\dot\Lambda}-8\pi \rho^{(\Lambda)} {\dot G}  }{8\pi G}= -\rho_c\frac{\dot G}{G}\not= 0,
\eea
where we made use of the EOS of the dark energy and the relation $\rho^{(\Lambda)}=\rho-\rho^{(m)}=\rho_c-\rho^{(m)}$. Multiplying eq. \eq{PLambda1} by the proper volume $\C{V}$ of the universe and comparing it by the first law of thermodynamics applied to the dark energy like in \cite{Mandal2021}, one reads off from eq. \eq{PLambda2} that the entropy $S^{(\Lambda)}$ of the dark energy changes, $T {\dot S}^{(\Lambda)}=\C{P}^{(\Lambda)}\C{V}$, where $T$ is the temperature of the dark energy. As a rule, the inequalities  $dG(k)/dk<0$  and ${\dot k}<0$ hold, implying that  ${\dot G} ={\dot k}(dG/dk)>0$, so that  $\C{P}^{(\Lambda)}<0$. Thus the entropy  $S^{(\Lambda)}$  should decrease in a realistic case. This does not contradict to the second law of thermodynamics, because the dark energy `in the auxiliary field $G(t)$' does not represent a closed thermodynamical system. Assuming that the matter and the dark energy are in thermodynamical equilibrium, the temperature $T$ can be estimated by making use of Stefan-Boltzmann law, $T =(\rho^{(m)}/\sigma)^{1/4}$ with the constant $\sigma$ determined by the number of independent degrees of freedom of matter.
Then we get the estimate
\bea\label{dSdt}
  {\dot S}^{(\Lambda)} &=& -\C{V} \sigma^{1/4}  \frac{\dot G}{G} \frac{\rho_c}{(\rho^{(m)})^{1/4} }.
\eea

The above listed equations incorporate implicitly the yet undetermined function $k=k(t)$, the $k$-to-$t$ conversion rule.  In the case of cosmological expansion it is expected that  the RG scale $k$ can be uniquely converted into the cosmological time $t$ in such a manner that strictly monotonically decreasing $k$ corresponds to continuously and strictly monotonically increasing $t$. Therefore, characteristic times $t_G< t_\Lambda$ can be assigned to the dynamical RG scales $k_\Lambda<k_G$. Let us turn now to the particular model characterized by the properties {\em (i)-(iv)} and described in terms of the cosmological variables $H(t)$, $\rho^{(m)}(t)=\rho^{(f)}(t)$ and the conversion rule $k=k(t)$.  For  $k<k_\Lambda$, i.e., for $t >t_\Lambda$ the couplings $G$ and $\Lambda$ are frozen at their values $G_0$ and $\Lambda_0$ taken at the scale $k_\Lambda$, the consistency condition \eq{consis} reduces to the law of local energy conservation of matter, i.e., to eq.   \eq{encon}, so that we have 2 independent equations, say  eqs. \eq{encon} and \eq{Fr1} for the 2 variables $H(t)$ and $\rho^{(f)}(t)$  for times $t>t_\Lambda$. Therefore, the classical evolution takes places for times $t>t_\Lambda$ with the initial conditions given at $t_\Lambda$. 
For times $0<t\le t_\Lambda$  quantum effects come into the play via the evolving gravitational couplings, we have one more variable, $k(t)$, as well as  one more equation, the reduced consistency condition \eq{rconsis}. We take the point of view of \cite{Babic2005} that for given RG flow $G(k)$ and $\Lambda(k)$ the reduced consistency condition \eq{rconsis} can be interpreted as an implicit equation for the determination of  the conversion rule $k=k(t)$.

\subsection{Scheme B}

Now we turn to the description of the model  with the properties {\em (i)-(iv)} in terms of the dimensionless cosmological variables proposed in \cite{Litim2011,Copeland1998,Copeland2006}. That model is the particular case of the models discussed in \cite{Litim2011}, although in our model the  scalar field $\varphi$ with the potential energy density $V(\varphi)$  is degenerated in the sense that it exhibits a single degree of freedom incorporated in the actual value of $\Lambda$. Namely, we can think of  the potential $V(\varphi)\equiv\rho^{(\Lambda)}$ being independent of the field variable $\varphi$ and the field variable to be  frozen at any constant value, since a field-independent potential has a continuum of minimum places. Below we apply the description given in \cite{Litim2011} to our particular model.  In such an approach the dimensionless variables are
\bea
&& x=\pm \sqrt{  \frac{{\dot\varphi}^2/2}{\rho_c}   }\equiv 0,~~y=\sqrt{  \frac{ V}{\rho_c} }=\frac{\sqrt{\Lambda}}{\sqrt{3} H}, \\
&&
z=\frac{ V_{,\varphi}(\varphi)}{8\pi G V(\varphi)}\equiv 0,~~\Omega^{(f)} =\frac{\rho^{(f)}}{\rho_c}
\eea
with $V_{,\varphi}=\partial V/\partial \varphi$, and also the quantity $\eta=V_{,\varphi\varphi}/(8\pi G V)=0$
vanishes identically. 
For later use we shall introduce the cosmological parameters $\Omega^{(\Lambda)}=\rho^{(\Lambda)}/\rho_c=y^2$ and $\Omega=\Omega^{(f)}+\Omega^{(\Lambda)}$. Instead of the cosmological time it is more convenient to use the dimensionless evolution parameter $N=-\ln a$ related to the time variable $t$ via the relation $Hdt=-dN$. Therefore, the $k$-to-$t$ conversion rule reappears through the yet unknown function $k=k(N)$. A general field-dependent potential as well as the position of its minimum exhibit  scale dependences, that should be invoked from the RG analysis of the scalar field coupled to gravity. Nevertheless, in our model the only scale dependence of the minimum value of the potential energy density  is already incorporated in the scale dependence of $\rho^{(\Lambda)}$, i.e. in that of the gravitational couplings $G$ and $\Lambda$. The RG scale dependences are characterized by the beta-functions given in \cite{Litim2011}
 \bea
 &&\eta_{\rm{RG}}=\frac{\partial \ln G}{\partial \ln k},~~\nu_{\rm{RG}} =\frac{\partial \ln \Lambda}{\partial \ln k}-\eta_{\rm{RG}},
\eea
which are independent of the field variable $\varphi$ in our case and the quantity $\sigma_{\rm{RG}}$ falls off trivially.
The law of the local energy conservation \eq{encon} applied to the barotropic fluid (c.f. eq. (22) in \cite{Litim2011}) rewrites as
\bea
  \label{dOmedNqi}
 \frac{d\Omega^{(f)}}{dN} &=& 3\gamma\Omega^{(f)}(1-\Omega^{(f)} )
 +\Omega \eta_{\rm{RG}} \frac{d\ln k}{dN}.
\eea
Eqs. (19) and (21) in \cite{Litim2011} for the evolution of $x$ and $z$ turn to the trivial identity $0=0$, while eq. (20) in \cite{Litim2011} rewrites as
\bea\label{dydNqi}
 \frac{dy}{dN}&=& y\biggl(- \frac{3}{2} \gamma  \Omega^{(f)}+ \hf (\eta_{\rm{RG}}+\nu_{\rm{RG}}) \frac{d\ln k}{dN}\biggr).
\eea
The Friedmann constraint \eq{Fr1}  rewrites as 
\bea\label{conFr}
0&=&1-y^2-\Omega^{(f)}
\eea
and the constraint on the beta-functions given by eq. (25) in \cite{Litim2011} yields
\bea\label{conqi}
0&=& \eta_{\rm{RG}}(k) + y^2 \nu_{\rm{RG}}(k).
\eea
In the particular model discussed here the constraint $\eq{conqi}$ is equivalent to the reduced consistency condition \eq{rconsis}, see Appendix \ref{app:equiv}. 
Now the dynamical equation (27) in \cite{Litim2011} gives
\bea\label{dlnkdN}
 \frac{d\ln k}{dN} &=& \frac{3}{2\alpha_{\rm{RG}}}\gamma\Omega^{(f)}
\eea
with
\bea\label{alphaRG}
\alpha_{\rm{RG}}&=&\hf \biggl\lbrack \eta_{\rm{RG}}+\nu_{\rm{RG}}) -\frac{d}{d\ln k}\ln \biggl(-\frac{\eta_{\rm{RG}}}{\nu_{\rm{RG}}}\biggr) \biggr\rbrack
\eea
in our case. In this description the quantum improvement is taken into account in the dynamical equations \eq{dOmedNqi} and  \eq{dydNqi} by the terms proportional to $d\ln k/dN$ and the dynamical eq. \eq{dlnkdN}. In the classical case the evolving scale $k=k(N)$ and the terms proportional to  $d\ln k/dN$  fall off. Due to the Friedmann constraint \eq{conFr} the classical model exhibits a single independent variable (e.g., $y$), while  the quantum improved one exhibits 2 independent variables, (e.g., $y$ and $k$). For the later discussion  it becomes useful the relation of the flow of the Hubble parameter to that of the RG scale $k$ given by eq. (28) in \cite{Litim2011}, which rewrites as
\bea\label{dlnHdN}
 \frac{d\ln H}{dN} &=& \alpha_{\rm{RG}}\frac{d\ln k}{dN} 
\eea
 in our case.   

\section{Analytical solution}

In this section we show that both Schemes A and B provide the same analytical result for the evolution of the discussed model universe along the entire time axis. First, the analytical solutions have been determined in the time intervals $[0,t_G]$, $[t_G,t_\Lambda]$, and $[t_\Lambda, \infty)$ separately and then these sectional solutions have been sewed continuously at the times $t_G$ and $t_\Lambda$ requiring the continuity of $k(t)$ at $t=t_G$ and that of $H(t)$ at $t=t_\Lambda$. The classical solution valid for $t_\Lambda\le t<\infty$ has been found in Scheme B, and the constant of integration in $H(t)$ settled by its continuity at $t=t_\Lambda$. The zero of the cosmological time is chosen when the expanding solution has its initial singularity.

Independently of the scheme used, the local energy conservation \eq{encon} for the barotropic fluid yields the relation
\bea\label{rhofM}
 \rho^{(f)}&=&\C{M} a^{-3\gamma},
\eea
where the (dimensionful) constant of integration $\C{M}$ is a free parameter, determining the matter content of the model universe. The relation \eq{rhofM} holds for all times $0<t<\infty$, so that $\C{M}$ remains the only free parameter after the solution in the various time intervals have been sewed continuously.

\subsection{Scheme A}\label{subsec:schemeA}

Applying the independent  eqs. \eq{encon}, \eq{Fr1}, and \eq{rconsis} to our model for $0<t\le t_\Lambda$, eqs. \eq{rconsis} and \eq{encon}  enable one to express the variables $\rho^{(f)}(t)$ and $H(t)$ in terms of the function  $k(t)$,
\bea\label{rhofk}
 \rho^{(f)}&=& -\frac{1}{8\pi} \frac{\dot\Lambda}{\dot G}
=\biggl\{ \begin{array}{ccc} 
  \frac{\lambda_*}{8\pi g_*} k^4 &{\rm{for}} & 0<t\le t_G,\cr
\frac{3\lambda_*}{8\pi G_0}k^2 &{\rm{for}}& t_G\le t\le t_\Lambda
\end{array},\\
\label{Hw}
H&=&- \frac{2}{3\gamma}\frac{\dot k}{k} \biggl\{\begin{array}{ccc}  2&{\rm{for}} & 0<t\le t_G,\cr  1
&{\rm{for}}& t_G\le t\le t_\Lambda  \end{array},
\eea
respectively, where we made use of the scale dependences given in eqs. \eq{Gk} and \eq{Lambdak}. Inserting the expressions \eq{rhofk} and \eq{Hw} into the Friedmann constraint \eq{Fr1}, one obtains a first-order   ordinary differential equation (ODE) for the function $k(t)$, the explicit form of which depends on the explicit scale dependence of the gravitational couplings given in eqs. \eq{Gk} and \eq{Lambdak}. In this manner we see that the dynamics contains a single additional constant of integration in both time intervals  $[0,t_G]$ and $[t_G,t_\Lambda]$ occurring when the ODE for the function $k(t)$ is integrated.

\subsubsection{Time interval $0<t\le t_G$}
The ODE for $k(t)$ takes the form
\bea\label{ODEkG}
 {\dot k} &=&   -\sqrt{\frac{3\lambda_*}{8} }\gamma k^2 ,
\eea
where we consider the case with negative square root the realistic one for the expanding universe,  since it yields increasing time $t$ with  decreasing RG scale $k$. The solution of eq. \eq{ODEkG} is given as
\bea\label{kt1}
k(t)&=&\sqrt{\frac{8}{3\lambda_*} }\frac{1}{\gamma t},
\eea
where we shifted the singularity of the solution to $t=0$ by the appropriate choice of the constant of integration.
Inserting this back into eqs. \eq{rhofk}, \eq{Hw}, \eq{Gk}, and \eq{Lambdak}, the time dependences of all interesting cosmological quantities can be  made explicit,
\bea
\label{Ht1}
 H(t)&=& \frac{4}{3\gamma t}\equiv\frac{h}{t},\\
\label{rhoft1}
 \rho^{(f)}(t)&=& \frac{8}{9\pi g_*\lambda_* \gamma^4 t^4}=\rho^{(\Lambda)}=\hf \rho_c,\\
\label{rhoct1}
\rho_{c}(t) &=& \frac{3H^2}{8\pi G} = \frac{16}{9\pi g_*\lambda_* \gamma^4 t^4},\\
\label{Omet1}
\Omega^{(f)}&=& \Omega^{(\Lambda)}=\hf.
\eea
The comparison of eqs. \eq{kt1} and \eq{Ht1} reveals that it holds  the simple proportionality 
\bea
k&=&  \sqrt{ \frac{3}{2\lambda_*} } H
\eea
in the time interval $0<t\le t_G$. 
The inverse proportionality of the Hubble parameter to the time $t$ leads to the time dependence of the scale parameter
\bea\label{at1} 
a=At^h,
\eea
 and the comparison of eq. \eq{rhofM} with \eq{rhoft1} gives
\bea
A  &=& \biggl(\frac{9\pi g_*\lambda_* \gamma^4}{8}\C{M}\biggr)^{\frac{1}{3\gamma} }.
\eea
We see that the usage of the reduced consistency condition \eq{rconsis} has lead to the naive inverse proportionality  $k\propto 1/t \propto H$ of the RG scale $k$ to the cosmological time $t$ in the interval $0<t\le t_G$.
 The cosmological evolution is unique in this interval for any given model characterized with a given value of the parameter $\C{M}$. Evaluating the right-hand side of the expression \eq{PLambda2},  it is straightforward to check that it holds the equality
\bea
 \C{P}^{(\Lambda)}&=&  -\frac{32}{9\pi g_*\lambda_* \gamma^4 t^5}
={\dot \rho}^{(\Lambda)}.
\eea
Furthermore, one gets from eq. \eq{dSdt} that ${\dot S}^{(\Lambda)} \propto - 1/t^4$, i.e., the rate of  the negative entropy production falls off  rapidly during the Planck era $0<t\le t_G$.

Since the solution is uniquely settled in the interval $[0,t_G]$, one is enabled to determine the characteristic time scale corresponding to the dynamical RG scale $k_G$ through the requirement $k(t_G-0)=k_G$, 
\bea\label{tG}
 t_G&=&\sqrt{\frac{8}{3\lambda_*} }\frac{1}{\gamma k_G}=\sqrt{ \frac{8}{9g_*\lambda_* \gamma^2} } m_{Pl}^{-1},
\eea
where we made use of eqs. \eq{kt1} and \eq{params}. Thus we obtained that $t_G\propto 1/k_G$ in the discussed model.

\subsubsection{Time interval $t_G\le t\le t_\Lambda$}

For later use we rewrite the RG scale dependences \eq{Gk} and  \eq{Lambdak} of the gravitational couplings as
\bea
&& G(k)= b\biggl( -k^2 +\hf E\biggr),~~~~\Lambda(k)= c ( k^4 +F)
\eea
in the interval $  k\in [k_\Lambda, k_G]$, where
\bea
 E&=& 2\biggl( \frac{G_0}{b} + k_\Lambda^2\biggr)=3k_G^2\biggl( 1+ \frac{2}{3}\frac{k_\Lambda^2}{k_G^2} \biggr),\\
F&=& \frac{\Lambda_0}{c} -k_\Lambda^4
=k_G^4 \biggl( \frac{ G_0\Lambda_0 }{3g_*\lambda_* } -\frac{k_\Lambda^4}{k_G^4} \biggr)\approx -8.4\times 10^{-120} k_G^4.\nn
&&
\eea
Taking again the negative square root, the dynamical equation for the RG scale $k=k(t)$ can now be recasted as
\bea\label{kODEII}
 {\dot k} &=& -\frac{3\gamma}{2} k D^{1/2},
\eea
with 
\bea\label{Dk}
D&=& \frac{\lambda_* G(k)k^2}{G_0} +\frac{\Lambda(k)}{3}
=-b_2u^2 +b_1 u +b_0,
\eea
where the variable $u=k^2$ and the coefficients
\bea\label{b0}
 b_0&=& \frac{\Lambda_0}{3} - \frac{\lambda_*G_0}{9g_*} k_\Lambda^4\approx -0.47\times 10^{-122} m_{\rm{Pl}}^2,\\
\label{b1}
 b_1&=&  \lambda_* \biggl( 1+ \frac{2G_0}{9g_*} k_\Lambda^2\biggr)\approx \lambda_* \lbrack 1+\ord{k_\Lambda^2/k_G^2}\rbrack ,\\
\label{b2}
 b_2&=& \frac{\lambda_* }{9g_*} m_{\rm{Pl}}^2
\eea
have been introduced. The quantity $D$ is
  definitively positive by construction in the interval $k\in [k_\Lambda,k_G]$. Multiplying both sides of eq. \eq{kODEII} by $2k$, one can rewrite it as  
\bea
 {\dot u} &=&-3\gamma u D^{1/2}(u).
\eea
The inverse function $t=t(u)$ is then given as
\bea\label{tIu}
 t-t_2&=& -\frac{1}{3\gamma} I(u)
\eea
with the constant of integration $t_2$ and the integral
\bea
 I(u)&=& \int \frac{du}{ u D^{1/2}(u)}.
\eea
Since the coefficient $b_0<0$ and the discriminant of the quadratic form $D(u)$ is $\Delta= -4b_0b_2-b_1^2\approx
-\lambda_*^2 \lbrack 1+\ord{k_\Lambda^2/k_G^2}\rbrack <0$, one gets (see item 2.266 in \cite{GR})
\bea
 I(u)&=& (-b_0)^{-1/2} {\rm{arc~tan~}} \C{I}(u)~~{\rm{with}}\nn
\C{I}(u)&=&\frac{2b_0+b_1u}{2 \lbrack -b_0D(u)\rbrack^{1/2} } 
\eea
and rewrites eq.   \eq{tIu} as
\bea
\C{I} (u)&=& \c{I}(t)\equiv\tan \biggl( - 3\gamma \sqrt{-b_0} (t-t_2)\biggr).
\eea
Since $b_1\approx \lambda_*$ with the  accuracy of 60 valuable digits, this yields the second-order algebraic equation
for $u$,
\bea\!\!\!\!\!
0&\approx & \lbrack \lambda_*^2 - 4b_0 b_2\C{I}^2(t)\rbrack u^2 
+ 4b_0 (\lambda_*u +b_0)\lbrack 1+ \C{I}^2(t)\rbrack
\eea
having  the roots
\bea\label{uroots}
 u_\pm &=&
\frac{ 2b_0}{ \lambda_*}\frac{ -1 \pm  \sqrt{ 1- \Xi(t) } }{\Xi(t)}
\eea
with
\bea\label{Xit}
 \Xi(t)&=& \frac{\lambda_*^2 -4b_0b_2 \C{I}^2(t)}{\lambda_*^2\lbrack 1+\C{I}^2(t)\rbrack } >0.
\eea
Below we  show that only the root $u_-$ is physical. 
Making use of eq. \eq{tG}, the constant of integration $t_2$ is settled by the relation $I(k_G^2)= -3\gamma(t_G-t_2)$ which yields
\bea\label{t2a}
 t_2&=& t_G+ \frac{1}{3\gamma \sqrt{-b_0}} {\rm{arc~tan~}} \C{I}(k_G^2).
\eea
The estimate of $\C{I}(k_G^2)$ with an accuracy of 60 valuable digits gives $10^{61}$. For a huge number $\C{I}$ it is a good approximation to take 
\bea\label{hugeestim}
 {\rm{arc~tan~}} \C{I} &\approx &\frac{\pi}{2}- \frac{1}{\C{I}}.
\eea
This estimate enables one to determine the second term on the right-hand side of eq. \eq{t2a} with an accuracy of the order $\ord{t_G/t_2}$  to get 
\bea\label{t2b}
 t_2&=&\frac{2}{3}t_G + \frac{\pi}{6\gamma\sqrt{-b_0}}
\eea
and hence
\bea
 \C{I}(t) &=& {\rm{cotan}}\biggl\lbrack 3\gamma\sqrt{-b_0} \biggl(t-\frac{2}{3}t_G\biggr)\biggr\rbrack.
\eea
As to the next the relation $\C{I}(k_\Lambda^2)=\C{I}(t_\Lambda)$ enables one to determine the time scale $t_\Lambda$ corresponding to the dynamical scale $k_\Lambda$. Here $\C{I}(k_\Lambda^2)\approx 2.6\times 10^{30}$ is also a huge number, so that one makes use  of the estimate \eq{hugeestim} once more to obtain
\bea\label{tLam}
 t_\Lambda&=&\frac{2}{3}t_G+ \frac{2}{3\gamma \sqrt{\lambda_*} k_\Lambda}.
\eea
This means that with an accuracy of 30 valuable digits $t_\Lambda$ is inversely proportional to $k_\Lambda$.
The knowledge of $t_\Lambda$ enables one to decide that only the root $u_-$ reproduces $k_\Lambda^2$ when the expression \eq{tLam} for  $t_\Lambda$ is inserted into the right-hand side of eq. \eq{uroots}. It is worthwhile mentioning that $\Xi(t_\Lambda)\approx 10^{-60}$ and $\Xi(t_G)\approx 10^{-122}$, so that $\Xi$ remains very small as compared to 1 in the entire interval $[t_G,t_\Lambda]$. This justifies the approximation $\sqrt{1-\Xi}\approx 1-\hf \Xi$. Therefore the root $u_-$ of \eq{uroots} provides with rather high accuracy that
\bea\label{ktappr}
k(t) &\approx & \sqrt{\frac{-4b_0}{\lambda_*\Xi} } \approx \sqrt{3} k_G ( 1+ B_H\B{t}^2)^{-1/2}
\eea
with  $B_H=\frac{81g_*\lambda_*\gamma^2}{4} m_{\rm{Pl}}^2$ and the shorthand notation $\B{t}=t-\frac{2}{3}t_G$.
It is straightforward to check that even with this estimated form of the conversion rule $k=k(t)$ it holds the equality  $k(t_G)=k_G$.
 
The comparison of eq. \eq{Hw} with eq. \eq{kODEII} yields 
\bea\label{HDk}
 H&=& \sqrt{D},
\eea
a rephrasing of the Friedmann constraint. Therefore, the relations \eq{rhofk}, 
\bea
&& \rho^{(\Lambda)}= \frac{c(k^4+F)}{8\pi b(-k^2+\hf E)}, ~~\rho_c=\frac{3D}{8\pi b(-k^2+\hf E)},
\eea
imply that $\rho_c=\rho^{(f)}+\rho^{(\Lambda)}$ and $\Omega^{(f)}+\Omega^{(\Lambda)}=1$ which is another form of the Friedmann constraint. Eqs. \eq{HDk} and \eq{Dk} reveal a nontrivial relationship between the RG  scale $k$ and the Hubble parameter $H$ in the time interval $t_G\le t\le t_\Lambda$ instead of the simple proportionality $k\propto H$  valid in the interval $0\le t \le t_G$. For later discussions, it is useful to rewrite \eq{kODEII} in the form
\bea\label{dlnkdttr}
 \frac{d\ln k}{dt} &=& -\frac{3\gamma}{2} H(k).
\eea
It is important to notice that both eqs. \eq{HDk} and \eq{dlnkdttr} are exact equations.

Since the expression \eq{ktappr} is an estimate, we may get somewhat different results for the function $H(t)$ when evaluating it in various ways,
\bea
&& H^{(1)}(t) = \sqrt{ D\bigl(k^2(t)\bigr) }~~{\rm{or}}~~
H^{(2)}(t)= -\frac{2{\dot k}}{3\gamma k},
\eea    
although $H^{(1)}(t)$ and $H^{(2)}(t)$ are identical for the exact form of the function $k(t)$. 
The nonvanishing of the quantity
\bea 
\delta_H(t)&=& \frac{H^{(2)} (t)-H^{(1)}(t) }{H^{(2)} (t)}
\eea
characterizes the accuracy of the usage of the estimate \eq{ktappr}. A straightforward but lengthy evaluation yields
\bea\label{H1t2}
H^{(1)}(t) &=& 3\sqrt{g_*\lambda_*} m_{\rm{Pl}}\biggl\lbrack 
-\frac{1}{(1+B_H\B{t}^2)^2} + \frac{1}{1+B_H\B{t}^2} \biggl( 1+\frac{2k_\Lambda^2}{3k_G^2}\biggr)
 + \frac{F}{9k_G^4}\biggr\rbrack^{1/2},\\
\label{H2t2}
H^{(2)}(t)&=& \frac{2}{3\gamma} \frac{ B_H\B{t} }{1+B_H\B{t}^2},\\
\label{deltaH}
\delta_H(t)&=& 1-\biggl\lbrack 1+ \frac{2k_\Lambda^2}{3k_G^2} + \frac{1}{B_H\B{t}^2} \frac{2k_\Lambda^2}{3k_G^2}+ \frac{ (1+B_H\B{t}^2)^2}{B_H\B{t}^2} \frac{F}{9k_G^4} \biggr\rbrack^{1/2} \equiv 1-A_H^{1/2}(t).
\eea
The numerical values of $\delta_H(t)$ are extremely small, e.g., one gets $\delta_H(t_G) \approx -\hf \frac{k_\Lambda^2}{k_G^2}\sim \ord{10^{-60}}$ and $\delta_H(t_\Lambda) \approx \frac{1}{6}\frac{ k_\Lambda^2}{k_G^2} -\frac{ \Lambda_0}{2\lambda_* k_\Lambda^2} \sim \ord{10^{-60}}$. Thus $H^{(1)}(t)$ and $H^{(2)}(t)$ provide
identical numerical results with the accuracy of 60 valuable digits. With the help of the estimate \eq{ktappr}, it is now straightforward to obtain  the time dependences of the other interesting cosmological quantities. In order not to violate  the Friedmann constraint by the time-dependent analytical expressions, we evaluate the critical density by making use of $H^{(1)}(t)$ given in eq. \eq{H1t2}. Then we obtain
\bea
\label{rhoft2}
\rho^{(f)}(t)&=&\frac{3\lambda_*m_{\rm{Pl}}^2 k^2}{8\pi}=\frac{B_H m_{\rm{Pl}}^2}{6\pi \gamma^2( 1 +B_H\B{t}^2 )},\\
\label{rhoct2} 
\rho_c(t)&=&\frac{3\lbrack H^{(1)}(t)\rbrack^2}{8\pi G(t)}= \frac{27g_*\lambda_* m_{\rm{Pl}}^2 A_H(t) }{8\pi b\lbrack -k^2(t) +\hf E\rbrack },\\
\label{Omeft2}
\Omega^{(f)}(t)&=& \frac{\rho^{(f)}(t)}{\rho_c(t)}
=\frac{1}{A_H(t)} \biggl\lbrack -\frac{2}{(1+B_H\B{t}^2)^2} + \frac{1}{1+B_H\B{t}^2} \biggl( 1 + \frac{2k_\Lambda^2}{3k_G^2} \biggr) \biggr\rbrack,\\
\label{OmeLamt2}
\Omega^{(\Lambda)}(t)&=&\frac{\Lambda(t)}{3\lbrack H^{(1)}(t)\rbrack^2} =
\frac{1}{A_H(t)} \biggl\lbrack \frac{1}{(1+B_H\B{t}^2)^2 }  + \frac{F}{9k_G^4} \biggr\rbrack.
\eea
Now it is trivial that adding eqs. \eq{Omeft2} and \eq{OmeLamt2} recapitulates the identity $\Omega^{(f)}+\Omega^{(\Lambda)}=1$. 
The time dependence of the Hubble parameter \eq{H2t2} implies the time dependence of the scale factor
\bea\label{at2}
 a(t)&=& C_a (1+B_H \B{t}^2)^{\frac{1}{3\gamma} },
\eea
where the constant of integration $C_a=\biggl( \frac{ 8\pi}{27g_*\lambda_* } \C{M} \biggr)^{\frac{1}{3\gamma}} m_{\rm{Pl}}^{-\frac{4}{3\gamma} } $ is set by the continuity of $a(t)$ at $t=t_G$.
The comparison of eqs. \eq{ktappr} and \eq{at2} yields the relation
\bea\label{ka2}
 k&=& \sqrt{ \frac{ 8\pi}{3\lambda_* m_{\rm{Pl}}^2 }\C{M} } a^{-\frac{3\gamma}{2}}.
\eea

 The quantity $-\C{P}^{(\Lambda)}$ characterizing the entropy production is
\bea
\!\!\!\! \C{P}^{(\Lambda)}(t)&=& -\rho_c\frac{\dot G}{G} =-\frac{ 81 g_* \gamma m_{\rm{Pl}}^4}{ 16\pi} \frac{k^2(t) D^{3/2}[ k^2(t)]  }{ (-k^2(t)+\hf E)^2},
\eea
where one has to insert the function $k(t)$ from eq. \eq{ktappr} into the right-hand side. Making use of eq. \eq{dSdt}, one can take the ratio of the rates of entropy production at the end $t_\Lambda$ and at the begining $t_G$ of the crossover era. A somewhat lengthy, but straightforward evaluation yields the estimate,
\bea
\!\!\!\!\!\! \frac{  {\dot S}^{(\Lambda)}|_{t=t_\Lambda}  }{ {\dot S}^{(\Lambda)}|_{t=t_G} } &\approx & \frac{\B{t}_\Lambda}{\B{t}_G} \biggl( \frac{k_\Lambda}{k_G}\biggr)^{7/2}\lbrack 1+ \ord{ k^2_\Lambda/k^2_G } \rbrack \approx 10^{-75}.
\eea
This means that the entropy production rate radically falls off during the crossover era.

It is straighforward to check that the cosmological quantities we are interested in are all continuous at $t=t_G$ at least with the accuracy of 60 valuable digits corresponding to the ratio $k_\Lambda^2/k_G^2$. Eq. \eq{tLam} shows that the characteristic time scale $t_\Lambda$ corresponding to the RG scale $k_\Lambda$ is equal to $\B{t}_\Lambda$ with the accuracy of $t_G/t_\Lambda\sim k_\Lambda/k_G \sim 10^{-30}$, i.e., with that of 30 valuable digits.

\subsubsection{Evolution for $t\ge t_\Lambda$}\label{sec:evlateA}

For times $t\ge t_\Lambda$   the well-known classical evolution takes place. The time-dependences of the cosmological variables $H(t)$, $\rho^{(f)}(t)$,  and $a(t)$ are given in eqs. \eq{Htcl}, \eq{rhoftcl}, and \eq{atcl}, respectively, rederived now in Scheme B.  The constant of integration $C_t$ can be settled by requiring the continuity of $H(t)$ at $t=t_\Lambda$. With the accuracy of 30 valuable digits  one gets $C_t\approx \frac{2}{3}t_G$. The explicit evaluations with the help of the formula \eq{ktappr} show again that the cosmological quantities considered here are continuous  at $t=t_\Lambda$ at least with the accuracy of 60 valuable digits. It has been checked analytically that $\rho^{(f)}(t)$ and $a(t)$ are also continuous with the same accuracy at $t=t_\Lambda$ for that choice of $C_t$. The constant of integration $C_t$ describes that the Big Bang singularity should be dated with $\frac{2}{3}t_G$ earlier due to quantum effects than it was extrapolated from the classical cosmology.

\subsection{Scheme B}\label{subsec:schemeB}

In this section, we show that the solution of the system of eqs. \eq{dOmedNqi}-\eq{dlnkdN} provides exactly the same evolution in the case of the particular model discussed by us as the one we obtained in the framework of Scheme A. For $0<t\le t_\Lambda$ we have 5 equations for 3 yet unknown functions: $\Omega^{(f)}(N),~y(N), ~k( N)$. In order to test that our interpretation of $\rho^{(\Lambda)}$ as a field-independent  potential energy density  of a homogeneous scalar field does not lead to any contradiction, we checked that  the constraints \eq{conFr}, \eq{conqi}, and  either eq. \eq{dOmedNqi} or eq. \eq{dlnkdN} represent systems of 3 independent equations. Namely, both eqs. \eq{dydNqi} and \eq{dlnkdN} can be derived from the system of equations   \eq{conFr}, \eq{conqi}, and \eq{dOmedNqi}.  In our particular model the functions $y(k)$ and $\Omega^{(f)}(k)$ are directly determined from the constraints \eq{conFr} and \eq{conqi},
\bea\label{y2Omef}
&& y^2(k)= -\frac{\eta_{\rm{RG}}(k)}{\nu_{\rm{RG}}(k)}, ~~\Omega^{(f)}(k)= 1+ \frac{\eta_{\rm{RG}}(k)}{\nu_{\rm{RG}}(k)},
\eea
and  the evolution $k(N)$ of the RG scale is governed by eq. \eq{dlnkdN}.

\subsubsection{Time interval $0<t\le t_G$}

Making use of the scale dependences given in eqs. \eq{Gk} and \eq{Lambdak} we find that the RG parameters are the  scale-independent constants $\eta_{\rm{RG}} =-2$, $\nu_{\rm{RG}} =4$, and therefore eq. \eq{alphaRG} yields $\alpha_{\rm{RG}}=1$ and the solution of the constraints \eq{conFr} and \eq{conqi} are the constants $y^2(=\Omega^{(\Lambda)})=\Omega^{(f)}=\hf$. Thus  the solution in the time interval $0<t\le t_G$  represents a cosmological fixed point, when the matter content is a mixture of the barotropic fluid and the dark energy with equal energy densities. The dynamical equation  \eq{dlnkdN} for the RG scale reduces to
\bea
 \frac{d\ln k}{dN} &\equiv&\frac{d\ln k}{d\ln a}=\frac{3\gamma}{2}\Omega^{(f)}=\frac{3\gamma}{4} ,
\eea
having the solution 
\bea\label{ka1}
 k&=&\C{K}_1a^{-3\gamma/4}
\eea
with the constant of integration $\C{K}_1$. 
Then one finds from eq. \eq{dlnHdN} that $\frac{d\ln H}{dN}=\frac{d\ln k}{dN}$, so that
\bea\label{Ha1}
 H&=&C_Hk=C_H\C{K}_1a^{- 3\gamma/4}
\eea
 with the constant of integration $C_H$. Eq. \eq{Ha1} with $H={\dot a}/a$ represents a first-order ODE for the scale factor $a(t)$ having the solution
\bea\label{at1L}
 a(t)&=& \biggl(\frac{ 3\gamma C_H\C{K}_1 }{4} t\biggr)^{\frac{4}{3\gamma} },
\eea
where the point of singularity has been shifted to  $t=0$ by the appropriate choice of the constant of integration.
Inserting the time dependence of the scale factor \eq{at1L} back into eqs. \eq{ka1} and \eq{Ha1}, one gets the explicit time dependence of the scale factor and that of the Hubble parameter,
\bea\label{kt1L}
 k(t)&=& \frac{4}{3\gamma C_H} \frac{1}{t},\\
\label{Ht1L}
H(t) &=& \frac{4}{3\gamma} \frac{1}{t}.
\eea
Then the knowledge of the $k$-to-$t$ conversion rule enables one to make explicit the time dependence of the quantities $G(t),~\Lambda(t),
~\rho_c(t)=\frac{3H^2}{8\pi G(t)}$, and $ \rho^{(f)}(t)=\rho^{(\Lambda)}(t)=\hf \rho_c(t)$. The comparison of $\rho^{(f)}(t)$ and $\rho^{(\Lambda)}(t) $ evaluated in that manner with the expression \eq{rhofM} and the definition $\rho^{(\Lambda)}=\frac{\Lambda}{8\pi G}$, respectively,  enables one to settle the yet undetermined constants of integration,
\bea
 && \C{K}_1= \biggl( \frac{16\pi g_*}{\lambda_*}\C{M} \biggr)^{1/4},~~~~C_H=\sqrt{\frac{2\lambda_*}{3} }.
\eea
In this manner we recapitulate all the results given in eqs. \eq{kt1}-\eq{at1} obtained in the framework of scheme A.
The additional point is that we could now identify this early evolution of the discussed model universe with a cosmological fixed point with evolving RG scale. 

\subsubsection{Time interval $t_G\le t\le t_\Lambda$}

Making use of the RG scale dependences of the gravitational couplings given in eqs. \eq{Gk} and \eq{Lambdak}, one establishes that the RG parameters are scale-dependent in the interval $t_G\le t\le t_\Lambda$,
\bea\label{params2}
\!\!\!\!\!\!&&\eta_{\rm{RG}} (k)=\frac{ -2k^2}{-k^2+\hf E},~~ \nu_{\rm{RG}} (k)=\frac{2k^2(-k^4+Ek^2 +F)}{(k^4+F)(-k^2+\hf E)},\nn
\!\!\!\!\!\!&&
\alpha_{\rm{RG}}(k) = 1+\frac{\eta_{\rm{RG}}(k)}{\nu_{\rm{RG}}(k)}=\Omega^{(f)}(k).
\eea 
Then eq. \eq{dlnkdN} yields the first-order ODE
\bea
 \frac{d\ln k}{dN}&=& \frac{3\gamma}{2}
\eea
for the evolution of the RG scale, having the solution
\bea\label{ka2L}
 k&=&\C{K}_2e^{\frac{3\gamma}{2}N} = \C{K}_2 a^{-\frac{3}{2}\gamma}
\eea
with the constant of integration $\C{K}_2$. Inserting the solution \eq{ka2L} into eq. \eq{dlnHdN}, one finds the first-order ODE
\bea
\frac{d\ln H}{dk} &=&\frac{\alpha_{\rm{RG}}(k)}{k}= 2k\frac{-k^2+\hf E}{-k^4+Ek^2+F},
\eea
having the solution
\bea\label{Hk2}
H(k) &=& D_H\sqrt{ -k^4+Ek^2+F}
\eea
with the constant of integration $D_H$. 
 With the help of the $k$-dependences of the Hubble parameter \eq{Hk2} and the RG parameters \eq{params2}, one is enabled to determine the $k$-dependences of the other important cosmological quantities,
\bea\label{rhock2}
\rho_c(k)&=&\frac{3H^2(k)}{8\pi G(k)} = \frac{3D_H^2}{8\pi b} \frac{-k^4 +Ek^2+F}{-k^2+\hf E},\\
\label{rhofk2}
\rho^{(f)}(k)&=& \rho_c(k)\Omega^{(f)}(k)=\frac{6D_H^2}{8\pi b} k^2=\frac{6D_H^2}{8\pi b}\C{K}_2^2a^{-3\gamma} ,\\
\label{OmeLamk2}
\Omega^{(\Lambda)}(k)&=&y^2(k)=\frac{ k^4 +F}{-k^4 +Ek^2+F},\\
\label{rhoLamk2}
\rho^{(\Lambda)}(k)&=&\rho_c(k)\Omega^{(\Lambda)} = \frac{3D_H^2}{8\pi b} \frac{k^4+F}{-k^2+\hf E}.\eea
The comparison of eqs. \eq{rhofk2} and \eq{rhoLamk2} with eqs. \eq{rhofM} and \eq{rhoLamdf}, respectively,
yields the expressions of the constants of integration,
\bea\label{DHcK2}
&&D_H=\sqrt{ \frac{\lambda_*}{9g_*m_{\rm{Pl}}^2} },
~~\C{K}_2=\sqrt{ \frac{ 8\pi }{\lambda_* m_{\rm{Pl}}^2} \C{M}}.
\eea 

Now we can compare our results \eq{HDk} and \eq{Hk2} for $H(k)$  obtained in the frameworks of Scheme A and Scheme B, respectively. Taking into account the definition \eq{Dk} of $D(k)$ and the explicit expressions \eq{b0}-\eq{b2} of the constants $b_0,~b_1$, and $b_2$, on the one hand, and    the explicit expression of the constant $D_H$
 given in \eq{DHcK2} and the relations $E=3b_1/c$, $F=3b_0/c$, and $3b_2/c=1$, on the other hand, one concludes that both schemes provide the same result for $H(k)$. Moreover, taking the time-derivative of both sides of eq. \eq{ka2L}, we just recover the first-order ODE \eq{dlnkdttr} found in scheme A. In Appendix \ref{app:equiv} we have shown that the constraint \eq{conqi} of scheme B is identical with the reduced consistency condition \eq{rconsis} of scheme A in the case of the particular model discussed by us. Then the dynamical problem of evolution to be solved turns out to be identical in both schemes. Thus we get the same time dependences of all cosmological quantities in both schemes in the time interval $t_G\le t\le t_\Lambda$.

\subsubsection{Evolution for $t\ge t_\Lambda$} 

For $t\ge t_\Lambda$ the RG scale is frozen at $k=k_\Lambda$ and the evolution becomes classical, described by 2 independent equations: for example, the classical versions of eq. \eq{dOmedNqi}, 
 \bea
  \label{dOmedNcl}
 \frac{d\Omega^{(f)}}{dN} &=& 3\gamma\Omega^{(f)}(1-\Omega^{(f)} ),
\eea
and the  Friedmann constraint \eq{conFr}. Furthermore, it holds the equation
\bea\label{dlnHdNcl}
 \frac{d\ln H}{dN} &=&\frac{3}{2}\gamma\Omega^{(f)},
\eea
which represents eq. (11) in \cite{Litim2011} applied to our particular model universe. 

 As a first step, one can determine the evolution of $\Omega^{(f)}$ vs. the parameter $N$ and express all cosmological quantities in terms of $\Omega^{(f)}$.
The first-order ODE \eq{dOmedNcl} can be solved in a straightforward manner by making use of the integral formula 2.103.4 in \cite{GR},
\bea\label{Omecl}
\Omega^{(f)} &=& ( 1+ C_\Omega e^{-3\gamma N})^{-1}= (1+ C_\Omega a^{3\gamma})^{-1}
\eea
with the constant of integration $C_\Omega$.
Then  one obtains trivially from the Friedmann constraint that
\bea\label{y2cl}
 y^2&=& \Omega^{(\Lambda)}=1-\Omega^{(f)}.
\eea
Inserting the function $\Omega^{(f)}=\Omega^{(f)}(N)$ given in eq. \eq{Omecl} into the right-hand side of eq. \eq{dlnHdNcl}, one obtains a first-order ODE for the $N$-dependence of the scale factor, having the solution
\bea
 H&=& \frac{ C_H C_\Omega^{1/2} }{\sqrt{ 1-\Omega^{(f)}}},
\eea
reexpressed in terms of  $\Omega^{(f)}(N)$, where $C_H$ is the constant of integration.
Then one finds by definition that
\bea\label{rhoccl}
\rho_c &=& \frac{\rho^{(\Lambda)}}{y^2} =\frac{\Lambda_0}{8\pi G_0 (1-\Omega^{(f)})},\\
\label{rhofcl}
\rho^{(f)}&=& \rho_c \Omega^{(f)} =\frac{\Lambda_0}{8\pi G_0} \frac{\Omega^{(f)}}{1-\Omega^{(f)}},
\eea
and from the second equality of eq. \eq{Omecl} that
\bea\label{aOmecl}
 a&=& \biggl(\frac{ 1-\Omega^{(f)} }{C_\Omega \Omega^{(f)} } \biggr)^{\frac{1}{3\gamma}}.
\eea

As to the next we determine the inverse function $t=t(\Omega^{(f)})$ by making use of the relation $dt=-dN/H$,
\bea\label{tvsOmecl} 
 t&=&-\int \frac{dN}{H} = - \frac{1}{3\gamma C_HC_\Omega^{1/2} } \int \frac{d\Omega^{(f)}}{\Omega^{(f)}
\sqrt{ 1- \Omega^{(f)}} }\nn
&=& -\frac{ 1}{3\gamma  C_HC_\Omega^{1/2} } \ln \frac{ 1-\sqrt{ 1-\Omega^{(f)} } }{
 1+ \sqrt{ 1-\Omega^{(f)} } }+C_t
\eea
with the constant of integration $C_t$. (Here we made use of the integral formula 2.224.5 in \cite{GR}.)
Now one has to settle the constants of integration. With the help of the relations \eq{rhofcl} and \eq{aOmecl} one expresses $\rho^{(f)}$ in terms of the scale factor $a$, and a comparison with the expression \eq{rhofM} yields then $C_\Omega= \frac{\Lambda_0}{8\pi G_0 \C{M}}$. The constant $C_H$ is easily settled by requiring the equality of $\rho_c=\frac{3H^2}{8\pi G_0}= \frac{C_H^2 C_\Omega}{1-\Omega^{(f)}}$ with the expression given in eq. \eq{rhoccl}. This implies $C^2_HC_\Omega=\Lambda_0/3$ and $C_H=\sqrt{\frac{8\pi G_0\Lambda_0 \C{M}}{3} }$.  Finally we invert the relation \eq{tvsOmecl} and obtain the well-known time dependences  obtained generally by solving the problem of evolution in terms of the traditional, dimensionful cosmological variables:
\bea\label{Htcl}
 H(t)&=& \sqrt{\frac{\Lambda_0}{3}} \biggl\lbrack \tanh \biggl( \frac{3\gamma}{2} \sqrt{ \frac{\Lambda_0}{3}} (t-C_t)\biggr)\biggr\rbrack^{-1},\\
\label{rhoftcl}
\rho^{(f)}(t)&=& \frac{\Lambda_0}{8\pi G_0}\biggl\lbrack \sinh \biggl( \frac{3\gamma}{2} \sqrt{ \frac{\Lambda_0}{3}} (t-C_t)\biggr)\biggr\rbrack^{-2},\\
\label{atcl}
a(t)&=& C_\Omega^{-\frac{1}{3\gamma}} \biggl\lbrack \sinh \biggl( \frac{3\gamma}{2} \sqrt{ \frac{\Lambda_0}{3}} (t-C_t)\biggr)\biggr\rbrack^{\frac{2}{3\gamma}}.
\eea
The determination of the constant of integration $C_t$ has been discussed at the end of  Section \ref{sec:evlateA}.

\subsection{Typical time scales}
In Sections \ref{subsec:schemeA} and \ref{subsec:schemeB} it was shown that both schemes provide the same evolution of the discussed particular model universe. This evolution is characterized by various time scales. The evolution following just the the Big Bang singularity is governed by the RG evolution of the gravitational couplings determined by the Reuter fixed point and ends up at $t_G=\xi^{\rm{UV}}/k_G$ (see eq. \eq{tG}) corresponding to the dynamical RG scale $k_G$. The inverse proportionality $t=\xi^{(UV)}/k$ is valid in the whole time interval $0<t\le t_G$ with the coefficient $\xi^{(UV)}=\sqrt{\frac{8}{3\lambda_*}} \frac{1}{\gamma t}$ (see eq. \eq{kt1}) in agreement with previous findings \cite{BonReu2002,Babic2005}. There exists another dynamical RG scale, $k_\Lambda$ 
such that for $k\le k_\Lambda$ the gravitational couplings keep their  constant values. The corresponding time scale $t_\Lambda$ is given by eq. \eq{tLam}. Since $t_G/t_\Lambda\sim\ord{10^{-30}}$, one finds that $t_\Lambda\approx \xi^{(IR)}/k_\Lambda$ with an accuracy of 30 valuable digits, where $\xi^{(IR)}= \frac{2}{3\gamma\sqrt{\lambda_*}}$. Our finding for $\xi^{(IR)} $ is rather similar to that found in \cite{BonannoReu2002} in the perturbative regime, $\xi^{(pert)}= \sqrt{\frac{2\omega}{3\nu} }\frac{1}{\gamma}$, where $\omega$ and $\nu$ are RG scheme-dependent constants in the perturbative expansion of the gravitational couplings close to the Gaussian fixed point.  
The intermediate era of the evolution is governed by the crossover RG scaling properties of the gravitational couplings showed up between the UV and deep IR scaling regions. The $k$-to-$t$ conversion rule is rather complicated in that era, being not at all a simple inverse proportionality,  but can be estimated with a good accuracy with the formula given in eq. \eq{ktappr}.

In addition to the time scales $t_G$ and $t_\Lambda$, there are 2 other  important time scales: {\em (i)} $t_d$, when the accelerating expansion turns into a decelerating one, and {\em (ii)} $t_a$, when the decelerating expansion turns again into an accelerating one. The time scale $t_a\sim \ord{\Lambda_0^{-1/2}}$  is a feature of the classical cosmology occurring due to the dominance of the nonvanishing cosmological constant $\Lambda_0$ in the late future. The scale $t_d$ is, however, affected by quantum effects.  The deceleration parameter 
\bea
 q&=& -\frac{a{\ddot a}}{{\dot a}^2} = -1-\frac{\dot H}{H^2}
\eea
can easily be calculated for the Planck and the crossover eras. One finds
\bea
 q&=&-1+\frac{3\gamma}{4}=\biggl\{ \begin{array}{ccc} 0 &{\rm{for}} & w=1/3\cr
                                                           -\frac{1}{4} & {\rm{for}} & w=0\end{array},
\eea 
for $0<t\le t_G$ and
\bea
q&=& -1 + \frac{3\gamma}{2}  -\frac{ 2}{ 27 g_*\lambda_* \gamma m_{Pl}^2\B{t}^2 }
\eea
for $t_G\le t\le t_\Lambda$. The deceleration parameter is continuous at $t=t_G$ and is a strictly monotonically increasing function of $t$ in the crossover era,
 taking the positive value
\bea
 q(t_\Lambda)&=&
  -1+\frac{3\gamma}{2} -\frac{ \gamma k_\Lambda^2}{2k_G^2}\nn
&=&
\biggl\{\begin{array}{ccc}1 -\frac{2}{3} \frac{k_\Lambda^2}{k_G^2}&{\rm{for}}&w=1/3 \cr \hf \biggl(1- \frac{k_\Lambda^2}{k_G^2}\biggr)&{\rm{for}}& w=0\end{array}\biggr\}
\eea
when the crossover era is ended. 
One can see that the value of the deceleration parameter is model-dependent: for radiation $(w=1/3)$ it is vanishing at $t=t_G$, so that in that case it holds $t_d=t_G$, but for dust ($w=0$) $q$ changes sign at $t_d>t_G$. The condition $q(t_d)=0$ yields that
\bea
 t_d&=& \frac{2}{3}t_G + \sqrt{ \frac{\gamma}{12} }t_G\biggl( -  1 + \frac{3\gamma}{2}\biggr)^{-1}\nn
&=&\biggl\{ \begin{array}{ccc} t_G &{\rm{for}} & w=1/3\cr
   \frac{2+\sqrt{3}}{3} t_G>t_G &{\rm{for}} & w=0\end{array} \biggr\}.
\eea

\section{Summary}

In this paper, we proposed simple, analytic formulas to describe the main features of the RG scale dependences of the gravitational couplings 
in the framework of 4-dimensional, 2-parameters EH gravity. These analytic formulas interpolate in a continuous manner between the Reuter fixed point governed UV scaling regime and the low-energy IR regime where the couplings take their observed classical values. The interpolation formulas were constructed to ensure the continuity of the energy density of matter at the border of the UV and crossover scaling regions in the framework of asymptotically safe cosmology. We applied the proposed interpolation formulas to an analytically solvable, homogeneous and isotropic, spatially flat model universe which  contains a classical barotropic fluid and the $\Lambda$-component identified here with dark energy. The local energy conservation of the barotropic fluid has been required separately. The problem of evolution was solved in two schemes. In Scheme A we followed the method used in \cite{Babic2005}, i.e.,  we worked in terms of the traditional, dimensionful cosmological variables and determined the conversion rule $k(t)$ between the RG scale $k$ and the cosmological time $t$ by means of the reduced consistency condition \eq{rconsis} of the quantum-improved Friedmann equations. In Scheme B we followed the method used in \cite{Litim2011}, i.e., we worked in terms of dimensionless cosmological variables and determined the conversion rule $k(t)$ from the Friedmann constraint induced constraint on the RG parameters \eq{conqi}. We have shown, that the constraints \eq{rconsis} and \eq{conqi} are equivalent in the case of the particular model. Therefore, both schemes provided just the same result for the evolution of the system.

In the framework of the particular model, explicit formulas are given for the characteristic time-scales $t_G$ and $t_\Lambda$ corresponding to the dynamical RG scales $k_G$ and $k_\Lambda$, respectively,  arising in the RG analysis of  EH gravity. We recovered the well-known result of quantum-improved cosmology, that in the Planck era it holds just the inverse proportionality $t\propto 1/k$. In the crossover era a nontrivial  relation has been obtained for the conversion rule $k(t)$, but it happens that at the end of the crossover era it holds the relation
$t_\Lambda\propto 1/k_\Lambda$ with an extremely high accuracy. The characteristic time scale $t_d$ was also discussed, when  the quantum effects driven accelerating expansion of the early universe goes over into a decelerating one. It has been shown that it holds $t_d\ge t_G$ and the value of $t_d$ depends on the matter content of the model. The determination of the solution in Scheme B enabled us to identify the cosmological fixed points, in the sense of the classification given in \cite{Litim2011}, from which the evolution starts and at which it ends up. As a by-product, it was found that the quantum effects result in the change of the entropy of the dark  energy.

\appendix

\section{Equivalence of the constraint \eq{conqi} and the reduced consistency condition \eq{rconsis}}\label{app:equiv}

In the particular model characterized with the properties {\em (i)-(iv)} the equivalence of the constraint \eq{conqi} on the RG parameters with the reduced consistency condition \eq{rconsis} is shown below. Reminding that the $t$-dependence of the gravitational couplings occurs implicitly via their $k$-dependences, \eq{rconsis} can be rewritten as
\bea
 - 0&=& 8\pi \rho^{(f)} \eta_{\rm{RG}} +\frac{\Lambda}{G} (\eta_{\rm{RG}}+\nu_{\rm{RG}}).
\eea
Dividing both sides of this equation by $8\pi \rho_c$, we get
\bea
 0&=& \Omega^{(f)} \eta_{\rm{RG}} + \Omega^{(\Lambda)}(\eta_{\rm{RG}}+\nu_{\rm{RG}})\nn
&=&
(  \Omega^{(f)} +y^2) \eta_{\rm{RG}} + y^2\nu_{\rm{RG}},
\eea
which is identical with the constraint \eq{conqi} since it holds  the relation $y^2=\Omega^{(\Lambda)}$  and the Friedmann constraint \eq{conFr} in our case.


\begin{thebibliography}{999}
\bibitem{fnote1} The position of the Reuter fixed point depends slightly on the details of the RG analysis and can be influenced by the matter content of the early  universe, but our considerations make use of  the existence of  the Reuter fixed point rather than its position.
\end{thebibliography}
\end{document}